# Concentration Fluctuations from Multinomial Probability Theory and the possible role in continuum Microkinetic Rate Theory

M.F. Francis
Los Alamos National Laboratories, Los Alamos, New Mexico, 87545

## Abstract

Recently, continuum Microkinetic Rate Theory (cMRT) has been advanced as a method of studying rates of systems, where deviations between observation and cMRT theory have been found, and it hypothesized that these deviations are linked either to oscillations or fluctuations. Multinomial probability theory (MPT) is used to derive analytical expressions for concentration fluctuations, giving the fluctuations as a function of average concentrations and sample sizes. MPT predictions of fluctuations in kinetically constrained systems are verified against kinetic Monte Carlo, and it analytically shown that MPT predicts canonical and grand canonical ensemble fluctuations. These fluctuation results are discussed in conjunction with cMRT deviations and it argued that fluctuations are not responsible for the deviations.

## Introduction

Assessing the rate of a systems level process is a general problem. When considering only an isolated reaction, methods such as transition state theory apply[1-3]; however, complex reaction networks require a systems level approach. Systems controlled by rates are referred to as 'kinetically constrained'. Perhaps the most commonly referred to kinetically constrained system is a flow reactor. In a flow reactor, material is infed at some rate, reacts inside the volume, and then exits. Reactor dynamics are controlled by the pumping rate, the allowable reactions, and the external temperature. Other common kinetically constrained systems are catalysis[4-8], corrosion[9-12], biochemistry[13,14], combustion, and power systems. Kinetically constrained systems are neither in the canonical ensemble, where the number is fixed, nor in the grand canonical ensembles, where the potential is fixed, falling into no thermodynamic ensemble[15,16] (Fig 1). With no formal model, understanding kinetically constrained systems has required extensive experiment[17-20], theoretical investigation[17-20], and simulation using molecular dynamics[21] or kinetic Monte Carlo[22,23,24,25].

An early attempt to model kinetically constrained systems was based on a continuum form of Microkinetic Rate Theory (cMRT)[24,25]. This cMRT describes the rate of change of concentrations, $\dot{\boldsymbol{\theta}}$, in terms of a set of independent first order reactions with rate constants, $\boldsymbol{k}^{(1)}$, second order reactions with rate constants, $\boldsymbol{k}^{(2)}$, a lattice dimension, $a$, gradient, $\partial/\partial x$, and curvature, $\partial^2/\partial x^2$, giving a total expression as

$$\dot{\boldsymbol{\theta}} = \boldsymbol{k}^{(1)} \cdot \boldsymbol{\theta} + \boldsymbol{k}^{(2)} \cdot \boldsymbol{\theta} \cdot \boldsymbol{\theta} + \frac{a^2}{2}\frac{\partial}{\partial x}\left(\boldsymbol{k}^{(2)} \cdot \frac{\partial}{\partial x}\boldsymbol{\theta}\right) \cdot \boldsymbol{\theta}. \qquad (1)$$

cMRT was largely successful, describing precisely first order reaction stationary states, transport processes, but contained some deviations for select second order system stationary states. The second order reaction deviations of cMRT were found to increase with lower concentration, lower diffusion rates, and in the absence of a reactivity gradient.

Two principal hypotheses for these deviations were offered: oscillations and fluctuations. An oscillation being a structured change in parameters, like a wave, and a fluctuation being the variation of a specific set of parameters. The oscillation hypothesis suggests that the concentration field may not be flat, but contain variations of second order or higher, resulting in contributions from the transport term of cMRT to the stationary state. The fluctuation hypothesis can be understood by imagining the variation in local concentration, and how this variation might influence a second order process; if a rate, $r$, is a second order function as, $r = k\theta^2$, and that this concentration fluctuates by some amount, $\delta$, then the rate associated with this fluctuation would be $r = \frac{1}{2}k(\theta + \delta)^2 + \frac{1}{2}k(\theta - \delta)^2 = k(\theta^2 + \delta^2)$. Both hypotheses are well posed; however, oscillations and fluctuations may be intertwined. This makes developing a well-defined model of either oscillations, or fluctuations, a potentially important step in solving the broad class of systems that are kinetically constrained.

Here, multinomial probability theory (MPT) is proposed as a basis for estimating concentration fluctuations. MPT and its fluctuations are derived assuming only that
(i) the number of arrangements of states may be counted using a combinatorial method, meaning for example that in chemistry when an A collides with a B, it does not matter which B, only that it is a B, and
(ii) that the observation of any one state does not affect the observation of another state, i.e. that observation probabilities are independent.

MPT is found to be predictive of fluctuations, but when cMRT deviations are reexamined, it is found that a straightforward second order fluctuation picture is unsuccessful, suggesting a reimagining of the possible role of fluctuations in a continuum rate theory, and that the missing continuum rate theory term is likely oscillations.

### Derivation of MPT, expectation value, variance, and covariance

**Multinomial Probability Theory (MPT)** is the probability distribution associated with drawing from a discrete number of states, $n_{states}$, each state with a defined and independent probability, $p_i$, giving the probability mass function, $f$, of drawing a specific number of each of the states, $x_i$, over the course of a defined number of samples, $n_{sample}$. The MPT probability mass function is first derived, then showed to integrate to one, providing an identity which is then used to derive MPT expected value, variance, and covariance.

**The MPT probability mass function**, $f_{MPT}$, may be derived by (i), counting the total number of states as a combinatorial and invoking the binomial theorem, and (ii), assuming the observation probabilities are independent, allowing one to assign probabilities to each combination.

The difference between a permutation and a combination is shown in Fig 2. In Fig 2 is shown three possible states which may be either red or blue. When all states are distinguishable, as by viewing the number, the system is a permutation with the total count given by the factorial of the number of states as $n_{sample}!$; when only some states are distinguishable, as with color, the system is a combination with the total count given by $n_{sample}!/n_{blue}!\,n_{red}!$. The MPT system is a combinatorial system, just as physical systems. It will be taken that the number of combinations of a two state system of size $n_{sample}$ is known by the binomial theorem to be $n_{sample}!/(x_1!\,x_2!)$. If the binomial counting theorem is invoked, and each count weighted according to its independent probability, the binomial probability mass function is given as

$$f_{binomial} = \frac{n_{sample}!}{x_1!x_2!} p_1^{x_1} p_2^{x_2}.$$

The number of combinations in a multinomial distribution may be derived by invoking the binomial theorem and by using inductive reasoning. If a sample size of $n_{sample}$ is given, and only two states are distinguishable, state 1 and a state greater than 1, labelled 1 and $> 1$, the combinatorial count is given as $n_{sample}!/(x_1!\,x_{>1}!)$. If states in the subsystem $> 1$ become distinguishable as 2 and $> 2$, the subsystem combinations are given by $x_{>1}!/(x_2!\,x_{>2}!)$. The total number of combinations is given by the product of the two systems as $[n_{sample}!/(x_1!\,x_{>1}!)][x_{>1}!/(x_2!\,x_{>2}!)] = n_{sample}!/(x_1!\,x_2!\,x_{>2}!)$. Inductively applying the same line of reasoning gives the total number of combinations in a multinomial distribution of sample size, $n_{sample}$, and a total number of allowable states, $n_{states}$, according to

$$\frac{n_{sample}!}{x_1!x_{>1}!} \prod_{i=2}^{n_{states}} \frac{x_{>i-1}!}{x_i!x_{>i}!} = \frac{n_{sample}!}{\prod_{i=1}^{n_{states}} x_i!}.$$

And because the probability of each individual combination is weighted by $\prod_{i=1}^{n_{states}} p_i^{x_i}$, the MPT probability mass function is given as

$$f_{MPT} = n_{sample}! \prod_{i=1}^{n_{states}} \frac{p_i^{x_i}}{x_i!}. \tag{2a}$$

$\sum f_{MPT} = 1$ can be shown using inductive reasoning and invoking the identity $\sum_{i=1}^{n_{states}} p_i = 1$. If the first sum of the probability mass function is considered, for a sample size of one, MPT gives $\sum_{i=1}^{n_{states}} f_{MPT}(n_{sample} = 1) = p_1 + p_2 + \cdots + p_{n_{states}}$ for a sample size of two $\sum_{i=1}^{n_{states}} f_{MPT}(n_{sample} = 2) = p_1(p_1 + p_2 + \cdots + p_{n_{states}}) + p_2(p_1 + p_2 + \cdots + p_{n_{states}}) + \cdots + p_{n_{states}}(p_1 + p_2 + \cdots + p_{n_{states}})$; and for a generalized number of samples

$$\sum f_{MPT} = \underbrace{\left( \sum_{i_{sample}=1}^{n_{states}} p_{i_{sample}} \cdots \underbrace{\left( \sum_{i_2=1}^{n_{states}} p_{i_2} \underbrace{\left( \sum_{i_1=1}^{n_{states}} p_{i_1} \right)}_{=1} \right)}_{=1} \right)}_{=1} = 1.$$

**The MPT expected value**, $E[x_i]$, can be shown using a few algebraic rearrangements and the identity $\sum f_{MPT} = 1$. The expected value of $x_i$ over $f_{MPT}$ is defined as

$$E[x_i] = \sum n_{sample}! \prod_{i=1}^{n_{states}} x_i \frac{p_i^{x_i}}{x_i!}, \tag{2b}$$

or

$$E[x_i] = \sum x_i \frac{n_{sample}!}{x_1!...x_i!...x_{n_{states}}!} p_1^{x_1} ... p_i^{x_i} ... p_{n_{states}}^{x_{n_{states}}}.$$

The equation is now rearranged by pulling out a factor of $n_{sample} p_i$ giving

$$E[x_i] = n_{sample} p_i \underbrace{\sum \frac{(n_{sample}-1)!}{x_1!...(x_i-1)!...x_{n_{states}}!} p_1^{x_1} ... p_i^{(x_i-1)} ... p_{n_{states}}^{x_{n_{states}}}}_{=1}.$$

The term in the sum has been labelled as equivalent to one, which can be shown by a variable change. Implementing the variable change of $M = n_{sample} - 1$ and $y = x_i - 1$, the term in the sum becomes an MPT subsystem, allowing the application of $\sum f_{MPT} = 1$, from which it follows that

$$E[x_i] = n_{sample} p_i. \tag{2c}$$

**The MPT variance**, $Var[x_i]$, can be shown using a similar number of algebraic rearrangements as with the expected value problem, and a similar application of the identity $\sum f_{MPT} = 1$. The variance of $x_i$ is given according to $Var[x_i] = E[(x_i - \mu_i)^2] = E[x_i^2] - E[x_i]^2$. $E[x_i]$ was previously determined, and now $E[x_i^2]$ is needed; alternatively, $E[x_i(x_i - 1)] = E[x_i^2] - E[x_i]$ can be calculated, and the variance determined according to $Var[x_i] = E[x_i(x_i - 1)] + E[x_i] - E[x_i]^2$, which will be easier. $E[x_i(x_i - 1)]$ is given as

$$E[x_i(x_i - 1)] = \sum x_i(x_i - 1) \frac{n_{sample}!}{x_1!...x_i!...x_{n_{states}}!} p_1^{x_1} ... p_i^{x_i} ... p_{n_{states}}^{x_{n_{states}}}.$$

The equation is now rearranged by pulling out a factor of $n_{sample}(n_{sample} - 1) p_i^2$ giving

$$E[x_i(x_i - 1)] = n_{sample}(n_{sample} - 1) p_i^2 \underbrace{\sum \frac{(n_{sample}-2)!}{x_1!...(x_i-2)!...x_{n_{states}}!} p_1^{x_1} ... p_i^{(x_i-2)} ... p_{n_{states}}^{x_{n_{states}}}}_{=1}.$$

As before, implementing variable changes of $M = n_{sample} - 2$, $y = x_i - 2$, the term in the sum becomes an MPT subsystem with $\sum f_{MPT} = 1$ from which it follows that $E[x_i(x_i - 1)] = n_{sample}(n_{sample} - 1) p_i^2$, giving

$$Var[x_i] = n_{sample} p_i (1 - p_i). \tag{2d}$$

**The MPT covariance**, $Covar[x_i, x_j]$, can be shown, again, using algebraic rearrangements and the application of $\sum f_{MPT} = 1$. The covariance is given according to $Covar[x_i, x_j] =$ $E[x_i x_j] - E[x_i]E[x_j]$. $E[x_i]$ was previously determined, and now need $E[x_i x_j]$ is needed. $E[x_i x_j]$ is defined as

$$E[x_i x_j] = \sum x_i x_j \frac{n_{sample}!}{x_1!...x_i!...x_j!...x_{n_{states}}!} p_1^{x_1} ... p_i^{x_i} ... p_j^{x_j} ... p_{n_{states}}^{x_{n_{states}}}.$$

The equation is now rearranged by pulling out a factor of $n_{sample}(n_{sample} - 1)p_i p_j$ giving

$$E[x_i x_j] = n_{sample}(n_{sample} - 1)p_i p_j \times$$
$$\underbrace{\sum \frac{(n_{sample}-2)!}{x_1!...(x_i-1)!...(x_j-1)!...x_{n_{states}}!} p_1^{x_1} ... p_i^{(x_i-1)} ... p_j^{(x_j-1)} ... p_{n_{states}}^{x_{n_{states}}}}_{=1}.$$

Implementing a variable change of $M = n_{sample} - 2$, $y_i = x_i - 1$, and $y_j = x_j - 1$, the term in the sum becomes an MPT subsystem with $\sum f_{MPT} = 1$ from which it follows that $E[x_i x_j] = n_{sample}(n_{sample} - 1)p_i p_j$, giving

$$Covar[x_i, x_j] = -n_{sample} p_i p_j. \tag{2e}$$

### Analogy to Physical Ensembles & Expectations

**Analogy.** Whether in the canonical, or grand canonical ensembles, the probability of each set of systems, $W\{x_i\}$, is given by a subsystem weight factor, $w_i$, the number of each subsystem, $x_i$, and the system size, $N_{sys}$, according to

$$W\{x_i\} = N_{sys}! \prod_i \frac{w_i^{x_i}}{x_i!}, \tag{3a}$$

where the differences between the canonical and grand canonical ensembles are the determinations of $w_i$ via the partition function or the grand partition function, respectively[15,16]. This probability factor is such that $\sum_i w_i = 1$. The average number of observations, $\langle x_i \rangle$, of some state, $i$, can be taken as a weighted average as

$$\langle x_i \rangle = \sum_{\{x_i\}} x_i W\{x_i\}. \tag{3b}$$

The functional form of the probabilities in the canonical and grand canonical ensembles is directly analogous to the MPT probability mass function, (3a) and (2a), and the method of determining the average number of states is directly analogous to the expected value problem, (3b) and (2b). MPT predicts canonical and grand canonical fluctuations.

Comparing the canonical and grand canonical ensembles with MPT encourages the analogy between an ensemble system of size, $N_{sys}$, and an MPT sample size, $n_{sample}$, viewing the observation of $n_{sample}$ as a sample of some macrostate and furthermore encourages the analogy between the probability of a state in an ensemble, $w_i$, and the probability of an observation in a sample, $p_i$.

Insight is to be born with this analogy, but also nuance important. The two assumptions used to derive MPT fluctuations, (i-ii), are all valid for the canonical and grand canonical ensembles; this means there will be true, systematic fluctuations within an ensemble by analogizing, $w_i \sim p_i$ and $N_{sys} \sim n_{sample}$, but furthermore that when an experimental observation is made, where the entire ensemble is not normally observed, the observed fluctuations will correspond to $n_{sample}$ and not $N_{sys}$.

This manuscript is concerned with addressing fluctuations in kinetically constrained systems. In kinetically constrained systems, there is yet no completely predictive model of the stationary state; however, it is known that the system will propagate according to combinatorial and not permutative counting, assumption (i), leaving the only question as to whether MPT applies or not as to whether observation probabilities can be approximated as independent, assumption (ii). Here, assumption (ii) is taken to be to be the case, and MPT predictions of fluctuations tested against a kinetically constrained system.

**Expectations**. In physical systems, the number of states, $x_i$, is not normally observed, but the concentration, $\theta_i = x_i/N_{sys}$. MPT interprets the fluctuations as variances, $\sigma_{ii}^{MPT\,2}$, and covariances, $\sigma_{ij}^{MPT\,2}$, of the average concentration, $\theta_i^{avg} = E[x_i]/N_{sys}$, allowing the conversion of MPT results 2c, d, and e into variations of averages giving

$$\sigma_{ii}^{MPT\,2} = \frac{\theta_i^{avg}(1-\theta_i^{avg})}{N_{sys}}, \tag{4a}$$

and for $i \neq j$ as

$$\sigma_{ij}^{MPT\,2} = -\frac{\theta_i^{avg}\theta_j^{avg}}{N_{sys}}, \tag{4b}$$

resulting in a covariance matrix, $\boldsymbol{\Sigma}$, which can be constructed as

$$\Sigma_{ij}^{MPT} = \sigma_{ij}^2. \tag{4c}$$

## Verification of MPT Fluctuations

**The kinetic Monte Carlo (kMC) method** is used as a numerical technique against which to verify MPT fluctuation predictions, as kMC is taken to be the gold standard of rate theory of physical systems. kMC importantly propagates the state structure stochastically, rather than deterministically, allowing a realistic description of fluctuations[22,23,26-28]. kMC propagates a system based on two equations[26], eqns 5. kMC randomly selects from reactions of index $l$, each with rate $r_l$, based on a probability, $P_l$, of

$$P_l = r_l/\sum_m r_m, \tag{5a}$$

and uses a random number between 0 and 1, $R$, to propagate time by $\Delta t$ according to

$$\Delta t = -ln(R)/\sum_m r_m. \tag{5b}$$

**Boundary conditions.** Three separate benchmark systems are simulated with the kMC. Each system is a square lattice with periodic boundary conditions. What distinguishes the systems are the dimensions. Three sets of lattices of lengths 10x10, 16x16, and 50x50 were used. The system size was varied to test any sampling size effect.

**Reaction Rates.** Within each system 16 sets of simulations are performed where rate constants are randomly varied between zero and one, allowing diverse sampling. For each case, three states are allowed: A, B, and C. In each case, all reactions are allowed: $A \rightarrow B, A \rightarrow C, B \rightarrow C$ and their reverse reactions, $B \rightarrow A, C \rightarrow A$, and $C \rightarrow B$. Discussion of the concentrations in the kMC system is given based on a coverage parameter, $\theta_i$, which is given as the number of the state $i$, $N_i$, divided by the total number of states, $N_{sys}$, according to $\theta_i = N_i/N_{sys}$. The initial conditions are defined at $t = 0$ as $(\theta_A, \theta_B, \theta_C) = (1,0,0)$; with time the kMC will come to a stationary state with each state having some long time concentration average, $\theta_i^{avg}$; the initial conditions do not affect the stationary state but are required simulation parameters.

**Determining $\sigma_{ii}^{kMC^2}$ and $\sigma_{ij}^{kMC^2}$.** kMC simulations were performed for $10^8$ time steps, and stationary state verification and analysis performed on the last $10^7$ time steps, $n_t=10^7$. The variances and covariances were then determined according to

$$\sigma_{ii}^{kMC^2} = E[\theta_i^2] - E[\theta_i]^2 = \left[\frac{1}{n_t}\sum_t \theta_i(t)^2\right] - \left[\frac{1}{n_t}\sum_t \theta_i(t)\right]^2,$$

and

$$\sigma_{ij}^{kMC^2} = E[\theta_i\theta_j] - E[\theta_i]E[\theta_j] = \left[\frac{1}{n_t}\sum_t \theta_i(t)\theta_j(t)\right] - \left[\frac{1}{n_t}\sum_t \theta_i(t)\right]\left[\frac{1}{n_t}\sum_t \theta_j(t)\right].$$

**Verification of variance & covariance.** Verifying the ability of MPT to predict fluctuations is best done using a parity plot – comparing MPT and kMC predictions over a large sample. The vertical axes of Figures 3 a and b are respectively the kMC sampled variances and covariances where the corresponding horizontal axes are the MPT prediction. The dashed lines across Figures 3 a and b are the parity lines representing MPT and kMC equivalence. The overlap of the MPT and kMC data with the parity line verifies that MPT is accurately able to capture the fluctuations. It is verified in the supplementary information (SI) that the functional form of these fluctuations is Gaussian, which has been treated elsewhere for the limiting cases of large samples of MPT systems.

## Discussion

**Fluctuations or oscillations in cMRT?** A model of fluctuations has been derived in detail in an effort to examine its role in a systems level rate theory, specifically cMRT. Deviations from cMRT have been observed for the model system of simple adsorption and associative desorption, a Langmuir system. In this Langmuir case, species from a gas phase, $B_{(gas)}$, bind to vacant surface sites, *, with a first order reaction, $B_{(gas)}+* \rightarrow B*$, undergo exchange diffusion, $B*+* \leftrightarrow *+B*$, until they collide and desorb, 2B*→2*. When cMRT was used to study this Langmuir system, agreement was broadly found, except for cases lower concentration and slower diffusion rates.

When considering cMRT predications of the stationary state, there are three possible stationary states of eqn 1 to consider: a homogeneous system, a system with fluctuations, and an inhomogeneous system with some unarticulated oscillations. If the correct solution is a homogeneous cMRT, the stationary state would predicted by

$$0 = \boldsymbol{k}^{(1)} \cdot \boldsymbol{\theta} + \boldsymbol{k}^{(2)} \cdot \boldsymbol{\theta} \cdot \boldsymbol{\theta}; \tag{6a}$$

if fluctuations were to play a role, fluctuations would enter with the square, and perhaps some scaling factor, $\alpha$,

$$0 = \boldsymbol{k}^{(1)} \cdot \boldsymbol{\theta} + \boldsymbol{k}^{(2)} \cdot \boldsymbol{\theta} \cdot \boldsymbol{\theta} + \alpha \boldsymbol{k}^{(2)} \cdot \boldsymbol{\Sigma}; \tag{6b}$$

and if oscillations were to play a role, there would be nonzero values of $\partial \boldsymbol{\theta}/\partial x$, $\partial^2 \boldsymbol{\theta}/\partial x^2$, or $\partial \boldsymbol{k}^{(2)}/\partial x$ resulting in a solution of the form

$$0 = \boldsymbol{k}^{(1)} \cdot \boldsymbol{\theta} + \boldsymbol{k}^{(2)} \cdot \boldsymbol{\theta} \cdot \boldsymbol{\theta} + \frac{a^2}{2} \frac{\partial}{\partial x}\left(\boldsymbol{k}^{(2)} \cdot \frac{\partial}{\partial x} \boldsymbol{\theta}\right) \cdot \boldsymbol{\theta}. \tag{6c}$$

The cases for this Langmuir system containing deviations is shown in Fig 4. The observed and MPT fluctuations are shown in Fig 4a, which show comparable agreement across the range of concentrations. Fluctuations are largest in magnitude about $\Theta_B = 0.5$ and symmetrically drop as one goes to $\Theta_B = 0.0$ or $\Theta_B = 1.0$, as predicted by MPT. The observed and homogeneous solution predictions are shown in Fig 4b. Agreement between cMRT and observations is excellent at higher coverages, but poor at lower coverages, meaning fluctuations and cMRT deviations do not correlate. When the cases of poorest agreement are examined more closely, Fig 5, it is seen that the deviation from the cMRT model is proportional to diffusion, diffusion which is not in the fluctuation model. The dependence of the deviation on diffusion, and not fluctuations, implies that the deviation is transport mediated.

The fact that the model deviations do not appear to be the result of fluctuations, and that no fluctuations appear in rates, suggests a closer look at the conjecture that the rate is explicitly dependent upon the square of concentrations. The probability of collision can be reimagined as a state counting problem - counting nearest neighbors. The nearest neighbor counting problem is a Bayesian probability question, the probability of finding B, given that A is found, $P(B|A)P(A)$. If the state counting is given as a Bayesian problem, the fluctuation would drop out, indicating cMRT, is not a Fokker-Planck problem, and that 6b does not apply.

**Limitations of MPT as a fluctuation model**. The main assumption in using MPT as a fluctuation model is that the probability of sampling a state is independent - $p_i$ does not influence $p_j$, assumption (ii).

Another assumption, which was implicit in the approach, is that the system was a homogeneous phase, wherein the total set of probabilities is defined for the whole system. A wide variety of phase separations and spatial inhomogeneities may exist. Shown in Fig 6 is an example phase separation, a spinodal, where a material will self-separate into two phases (dark blue and turquoise in Fig 6). In the case of phase separation, the MPT fluctuation model applies for the separate phases, but not the system as whole, with the same reasoning extending to any inhomogeneous system.

## Conclusions

    Here has been derived a model of concentration fluctuations based on Multinomial Probability Theory (MPT). This model assumes, (i), combinatorial and not permutative counting of ensembles, and (ii), that the probability of drawing each state is independent. This MPT based model of concentration fluctuations is a function only of the average concentration, eqns 4. It has been argued that the model will apply to canonical, grand canonical, and kinetically constrained systems. The application of MPT to the canonical and grand canonical ensembles has been argued analytically, and the application of MPT to kinetically constrained systems verified against kMC. The model is limited to systems where the concentration profile is well defined.
    This work was motivated by the recent advances in continuum rate modelling, specifically continuum Microkinetic Rate Theory (cMRT), where it was conjectured that observation-cMRT deviations could be the result of either fluctuations or oscillations. It has been argued that the cMRT deviations are not the result of fluctuations, but connected to transport.

**Competing financial interests.** The author declares no competing financial interests.

**Author Contributions.** This work was performed solely by MFF.

**Acknowledgements.** This work was performed at Los Alamos National Laboratory, operated by Los Alamos National Security Administration, LLC, for the National Nuclear Security Administration of the US Department of Energy under contract DE-AC52-06NA25396. Support for this work was provided by the Director's Office under the Director's Fellowship program.

MFF would like to EF Holby for his support and the review of this paper; and to thank ES Martinez and AF Voter for suggesting a connection to Fokker-Planck.

**Materials and Correspondence.** Los Alamos National Laboratories, Los Alamos, New Mexico, 87545. mff7d@virginia.edu.

**Figures**

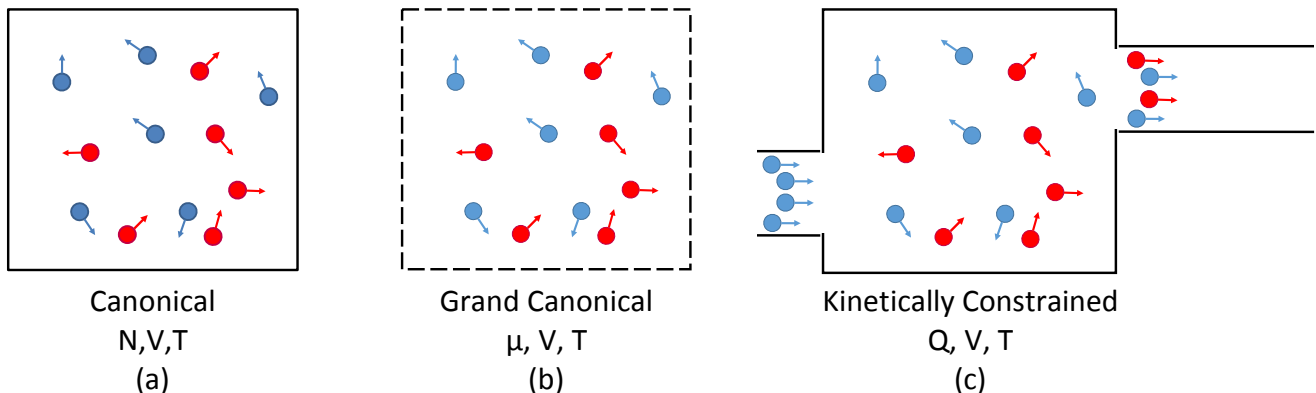

Fig 1. **A schematic representation of the canonical ensemble, grand canonical ensemble, and kinetically constrained systems**. Each system is centered about a reactive volume. Red and blue spheres are depicted to illustrate the possible reaction between these two states, where the added arrows are there to indicate a velocity. In the canonical ensemble, (a), the number of particles, N, is fixed, contained at some volume, V, and allowed to exchange energy with the outside world at a fixed temperature. The solid boundary in (a) is present to indicate an impermeable boundary. In the grand canonical ensemble, (b), particles are allowed to exchange with the outside world coming to a fixed chemical potential, μ; the system is at a fixed volume, V; and energy is allowed to exchange with the outside world at a fixed temperature, T. The dashed boundary in (b) is present to indicate a permeable boundary. Kinetically constrained systems, (c), are neither, controlled by a fixed flow rate, Q, fixed volume, and external temperature. The input and output feed in (c) are present to indicate a flow constraint. This image taken from reference[24,25].

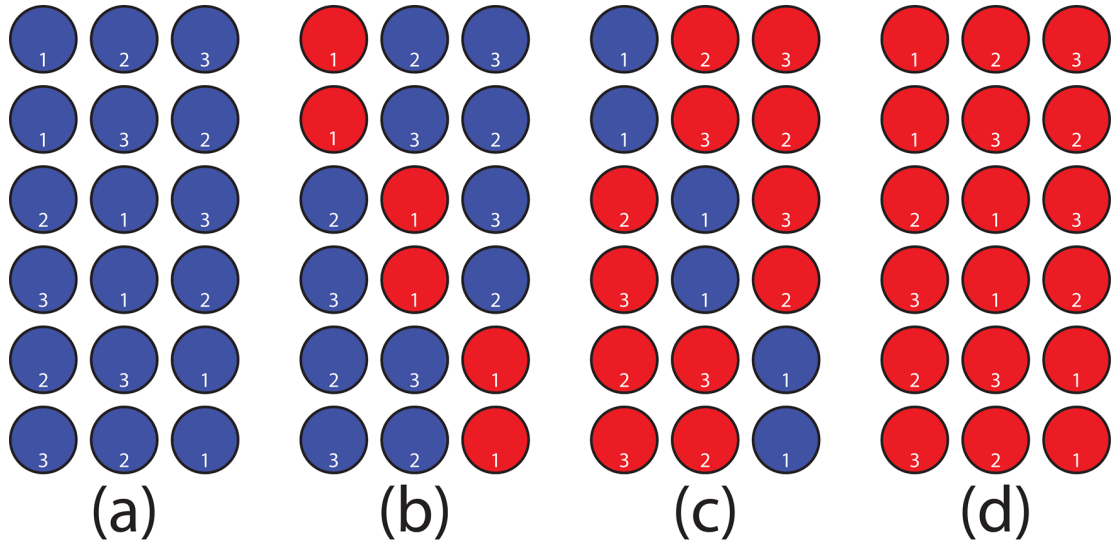

**Fig 2. An illustration of the difference between permutations and combinations, with permutation illustrated by number and combination illustrated by color.** (a), the state configuration is ($n_{sample}$=3, $n_{blue}$=3, $n_{red}$=0) where the number of permutations is given by $n_{sample}!=3!=6$ and the number of combinations is given by $n_{sample}!/(n_{blue}!n_{red}!)=3!/(3!0!)=1$. (b), the state configuration is ($n_{sample}$=3, $n_{blue}$=2, $n_{red}$=1) where the number of permutations is given by $n_{sample}!=3!=6$ and the number of combinations is given by $n_{sample}!/(n_{blue}!n_{red}!)=3!/(2!1!)=3$. (c), the state configuration is ($n_{sample}$=3, $n_{blue}$=1, $n_{red}$=2) where the number of permutations is given by $n_{sample}!=3!=6$ and the number of combinations is given by $n_{sample}!/(n_{blue}!n_{red}!)=3!/(1!2!)=3$. (d), the state configuration is ($n_{sample}$=3, $n_{blue}$=0, $n_{red}$=3) where the number of permutations is given by $n_{sample}!=3!=6$ and the number of combinations is given by $n_{sample}!/(n_{blue}!n_{red}!)=3!/(0!3!)=1$.

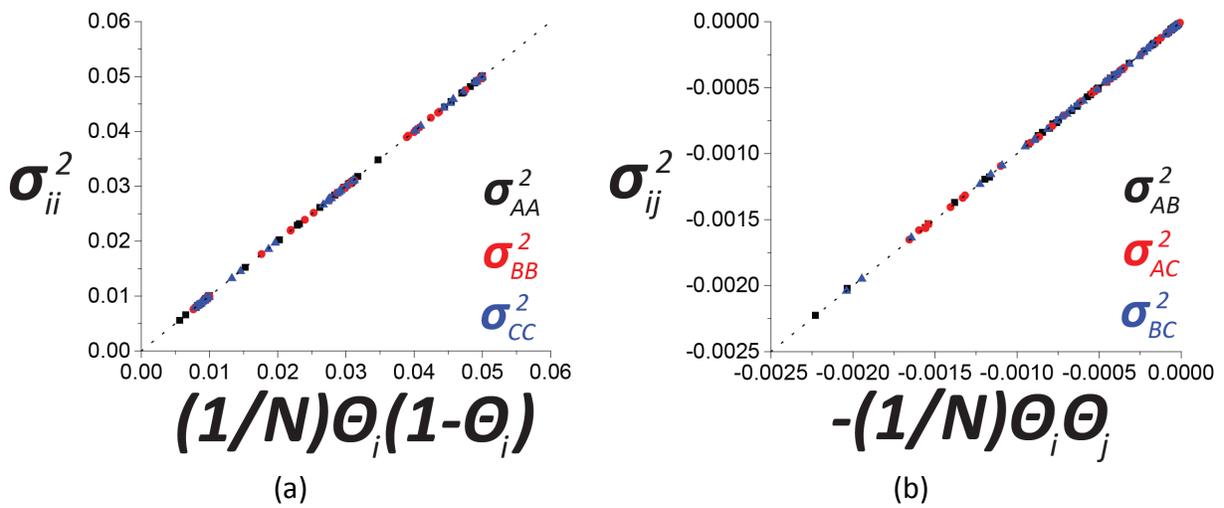

**Figure 3. A demonstration of MPT's ability to capture kMC predictions of, (a), variance and, (b), covariance.** (a) and (b) are graphical representations of the MPT and kMC variances and covariances with the data shown in Tables S2 and S3.

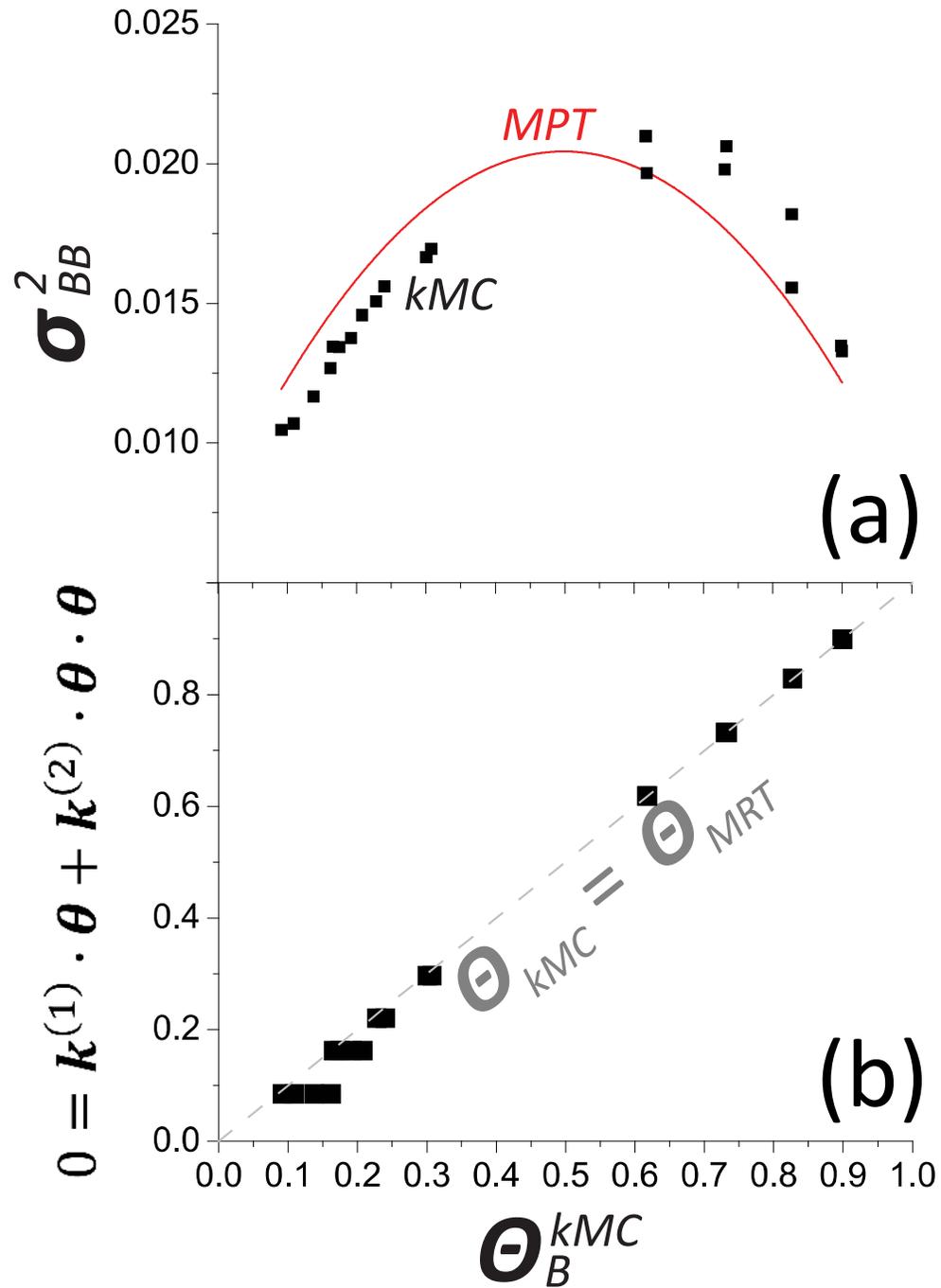

**Figure 4. Comparison of kMC and MRT stationary state properties for the cases including deviations from the model.** (a), comparison of variance as a function of stationary state showing the MPT model prediction in red and the kMC numerical results in black. (b), MRT prediction as a function of kMC numerical result for the stationary states.

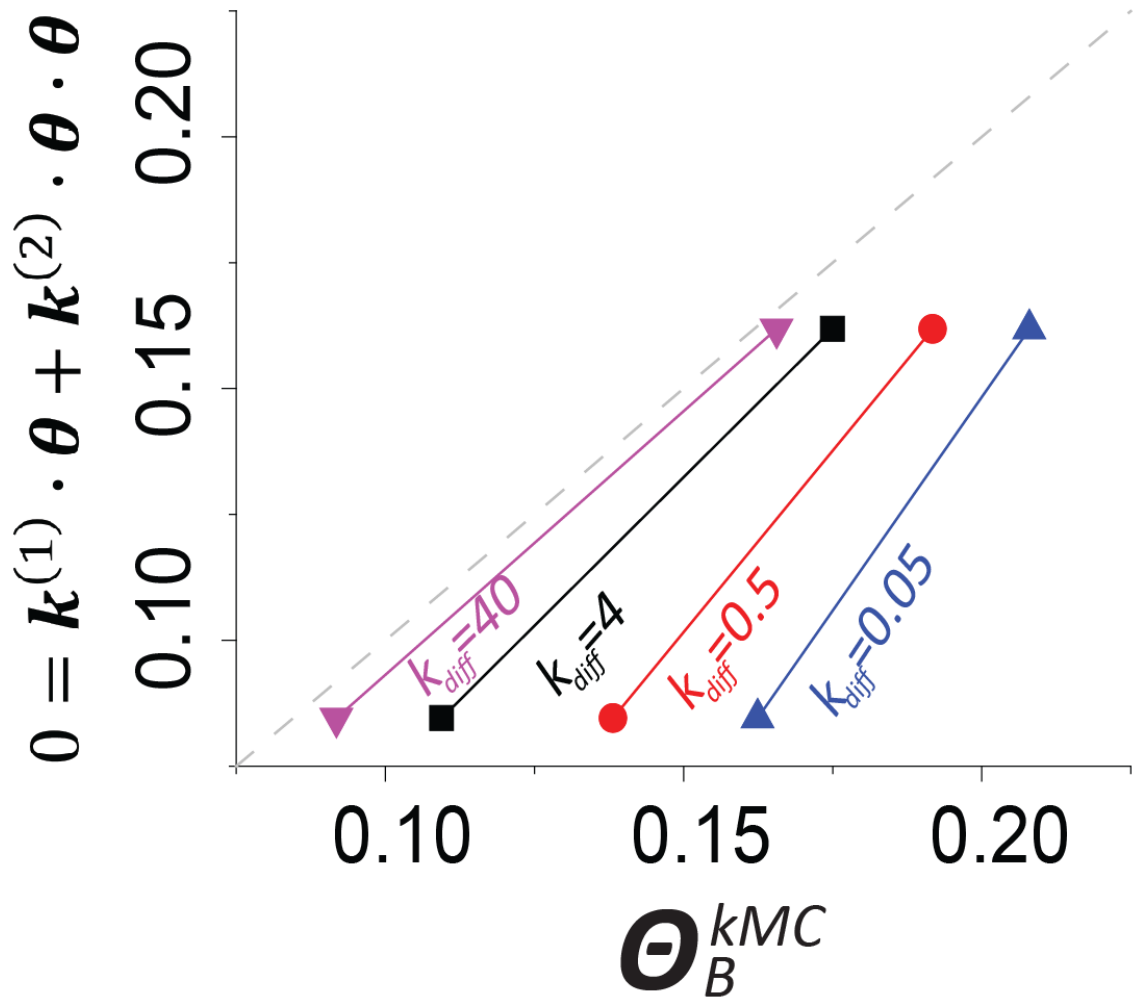

**Figure 5. Comparison of kMC and MRT stationary state properties, revealing diffusion to be controlling.** This is a zoom in of part of Figure 4.

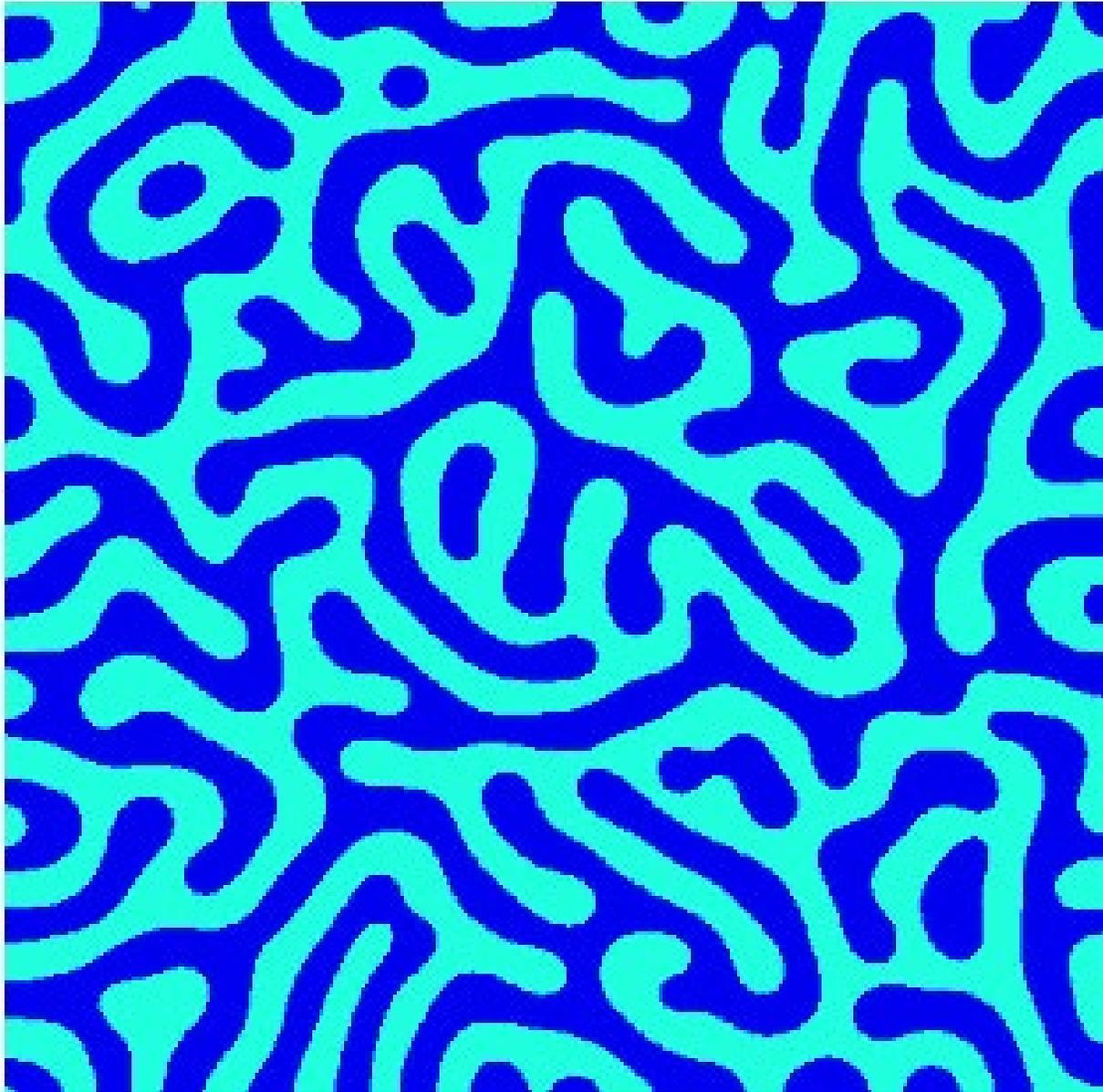

**Figure 6. Microstructures of a phase separation obtained from the Cahn-Hilliard-Cook model[29-31]**. The dark blue and light blue are two separate phases. A schematic description of different regions which may have different fluctuations.

## Supplementary Information

Shown in table S1 are the kMC parameters randomly selected for and used for the MPT verification in the body of the manuscript. The resulting kMC observations of stationary state is shown in table S2 along with variances and covariances in table S3.

The MPT predicted variance and covariance are used as parameters to a Gaussian model. These MPT predictions are compared to kMC simulations, shown in Fig S1, verifying the Gaussian form of the fluctuations. It is important to note that the MPT prediction of variance is for a standard Gaussian, and that if a non-standard Gaussian is used, the variance will shift by the linear factor to scale back to the standard form.

| Simulation Number | $N_{sys}$ | $k_{rxn}^{A \to B}$ | $k_{rxn}^{B \to A}$ | $k_{rxn}^{A \to C}$ | $k_{rxn}^{C \to A}$ | $k_{rxn}^{B \to C}$ | $k_{rxn}^{C \to B}$ |
|---|---|---|---|---|---|---|---|
| 1 | 100 | 0.102574 | 0.645494 | 0.843196 | 0.41223 | 0.511407 | 0.384304 |
| 2 | 100 | 0.651832 | 0.421919 | 0.0372046 | 0.79781 | 0.0380532 | 0.108896 |
| 3 | 100 | 0.613276 | 0.456051 | 0.087582 | 0.717818 | 0.218361 | 0.812635 |
| 4 | 100 | 0.456732 | 0.684033 | 0.195886 | 0.914937 | 0.659308 | 0.445055 |
| 5 | 100 | 0.440414 | 0.608314 | 0.914689 | 0.686433 | 0.782511 | 0.292201 |
| 6 | 100 | 0.87062 | 0.135397 | 0.789007 | 0.300414 | 0.516114 | 0.547627 |
| 7 | 100 | 0.900418 | 0.584832 | 0.952246 | 0.990299 | 0.322337 | 0.382642 |
| 8 | 100 | 0.431233 | 0.470224 | 0.603575 | 0.821936 | 0.887284 | 0.188041 |
| 9 | 100 | 0.699919 | 0.383928 | 0.278668 | 0.937976 | 0.383952 | 0.298865 |
| 10 | 100 | 0.829007 | 0.213554 | 0.378389 | 0.1609 | 0.437321 | 0.899987 |
| 11 | 100 | 0.729522 | 0.688995 | 0.0315202 | 0.547634 | 0.864919 | 0.989409 |
| 12 | 100 | 0.412546 | 0.941655 | 0.448052 | 0.770389 | 0.997377 | 0.931954 |
| 13 | 100 | 0.380019 | 0.53553 | 0.201622 | 0.0889054 | 0.850243 | 0.357466 |
| 14 | 100 | 0.0382846 | 0.636134 | 0.788824 | 0.172777 | 0.422213 | 0.57411 |
| 15 | 100 | 0.721077 | 0.952499 | 0.00178384 | 0.439105 | 0.934631 | 0.113399 |
| 16 | 100 | 0.834618 | 0.144919 | 0.168627 | 0.0335455 | 0.523613 | 0.692554 |
| 1 | 256 | 0.982006 | 0.843382 | 0.32429 | 0.965038 | 0.0205456 | 0.445286 |
| 2 | 256 | 0.729484 | 0.833111 | 0.164968 | 0.509313 | 0.889746 | 0.75083 |
| 3 | 256 | 0.213156 | 0.826441 | 0.472904 | 0.342124 | 0.193221 | 0.133167 |
| 4 | 256 | 0.953912 | 0.112119 | 0.839862 | 0.134555 | 0.485616 | 0.780623 |
| 5 | 256 | 0.172085 | 0.342865 | 0.874006 | 0.140143 | 0.459368 | 0.332817 |
| 6 | 256 | 0.588296 | 0.657108 | 0.122148 | 0.142694 | 0.431678 | 0.622145 |
| 7 | 256 | 0.876964 | 0.787113 | 0.872178 | 0.761923 | 0.710075 | 0.296427 |
| 8 | 256 | 0.460904 | 0.76933 | 0.97508 | 0.168301 | 0.287345 | 0.111455 |
| 9 | 256 | 0.420512 | 0.951317 | 0.122213 | 0.60783 | 0.532632 | 0.0858723 |
| 10 | 256 | 0.313255 | 0.959878 | 0.779915 | 0.239283 | 0.65612 | 0.100021 |
| 11 | 256 | 0.988937 | 0.222169 | 0.827579 | 0.259257 | 0.646045 | 0.364863 |
| 12 | 256 | 0.26819 | 0.237041 | 0.136221 | 0.846296 | 0.0553036 | 0.998964 |
| 13 | 256 | 0.35173 | 0.974044 | 0.3072 | 0.594545 | 0.12106 | 0.142345 |
| 14 | 256 | 0.232515 | 0.264558 | 0.0150576 | 0.547689 | 0.183832 | 0.872388 |
| 15 | 256 | 0.269704 | 0.652303 | 0.860944 | 0.517064 | 0.229583 | 0.891586 |
| 16 | 256 | 0.329604 | 0.719165 | 0.506002 | 0.152046 | 0.551773 | 0.978422 |
| 1 | 2500 | 0.497791 | 0.438188 | 0.0695434 | 0.12843 | 0.379287 | 0.739206 |
| 2 | 2500 | 0.238616 | 0.960725 | 0.334395 | 0.279784 | 0.721753 | 0.527619 |
| 3 | 2500 | 0.256975 | 0.26003 | 0.655425 | 0.305549 | 0.145088 | 0.556246 |
| 4 | 2500 | 0.0946682 | 0.650399 | 0.610365 | 0.571332 | 0.393757 | 0.688417 |
| 5 | 2500 | 0.131265 | 0.578722 | 0.519033 | 0.461369 | 0.0491759 | 0.531358 |
| 6 | 2500 | 0.82087 | 0.600901 | 0.959161 | 0.338447 | 0.259058 | 0.729332 |
| 7 | 2500 | 0.998264 | 0.0637266 | 0.577063 | 0.298817 | 0.958989 | 0.343511 |
| 8 | 2500 | 0.486608 | 0.998936 | 0.555791 | 0.700879 | 0.746639 | 0.304484 |
| 9 | 2500 | 0.302885 | 0.91485 | 0.795547 | 0.189304 | 0.953284 | 0.486181 |
| 10 | 2500 | 0.6417 | 0.00521446 | 0.32057 | 0.369746 | 0.220422 | 0.466584 |
| 11 | 2500 | 0.75178 | 0.425744 | 0.190615 | 0.449673 | 0.352681 | 0.764192 |
| 12 | 2500 | 0.0820129 | 0.341255 | 0.447937 | 0.406926 | 0.145739 | 0.640072 |
| 13 | 2500 | 0.48925 | 0.195863 | 0.893534 | 0.640173 | 0.488186 | 0.896742 |
| 14 | 2500 | 0.79267 | 0.69229 | 0.943058 | 0.896341 | 0.70752 | 0.881594 |
| 15 | 2500 | 0.193701 | 0.202164 | 0.538041 | 0.758463 | 0.198915 | 0.571909 |
| 16 | 2500 | 0.665499 | 0.762525 | 0.510242 | 0.862924 | 0.0912434 | 0.212198 |

**Table S1**. A list of the cell sizes and reaction constants used in the benchmark kMC simulations. Data is given to the same significant figures as was used in simulation for possible reproducibility.

| Simulation Number | $N_{sys}$ | $\theta_A^{avg}$ | $\theta_B^{avg}$ | $\theta_C^{avg}$ |
|---|---|---|---|---|
| 1 | 100 | 0.33619 | 0.18869 | 0.47512 |
| 2 | 100 | 0.39382 | 0.56597 | 0.04021 |
| 3 | 100 | 0.41518 | 0.49015 | 0.09466 |
| 4 | 100 | 0.54464 | 0.25351 | 0.20186 |
| 5 | 100 | 0.32893 | 0.2029 | 0.46817 |
| 6 | 100 | 0.11355 | 0.48596 | 0.40048 |
| 7 | 100 | 0.29061 | 0.41167 | 0.29772 |
| 8 | 100 | 0.40756 | 0.18607 | 0.40637 |
| 9 | 100 | 0.36878 | 0.41797 | 0.21325 |
| 10 | 100 | 0.14013 | 0.57279 | 0.28708 |
| 11 | 100 | 0.4535 | 0.34388 | 0.20262 |
| 12 | 100 | 0.49545 | 0.23585 | 0.2687 |
| 13 | 100 | 0.27123 | 0.21061 | 0.51816 |
| 14 | 100 | 0.29123 | 0.25658 | 0.45219 |
| 15 | 100 | 0.46513 | 0.19998 | 0.33489 |
| 16 | 100 | 0.08831 | 0.51777 | 0.39392 |
| 1 | 256 | 0.39827 | 0.50276 | 0.09896 |
| 2 | 256 | 0.43119 | 0.30029 | 0.26852 |
| 3 | 256 | 0.40023 | 0.14372 | 0.45605 |
| 4 | 256 | 0.06343 | 0.57383 | 0.36274 |
| 5 | 256 | 0.16471 | 0.26993 | 0.56536 |
| 6 | 256 | 0.38169 | 0.35625 | 0.26205 |
| 7 | 256 | 0.30629 | 0.26447 | 0.42924 |
| 8 | 256 | 0.15948 | 0.14321 | 0.69731 |
| 9 | 256 | 0.58105 | 0.17934 | 0.23961 |
| 10 | 256 | 0.22697 | 0.08704 | 0.68599 |
| 11 | 256 | 0.11913 | 0.35572 | 0.52516 |
| 12 | 256 | 0.41379 | 0.53886 | 0.04735 |
| 13 | 256 | 0.5364 | 0.20584 | 0.25776 |
| 14 | 256 | 0.54746 | 0.39509 | 0.05745 |
| 15 | 256 | 0.3453 | 0.38119 | 0.27351 |
| 16 | 256 | 0.34617 | 0.33515 | 0.31868 |
| 1 | 2500 | 0.37054 | 0.41743 | 0.21203 |
| 2 | 2500 | 0.46939 | 0.17744 | 0.35316 |
| 3 | 2500 | 0.23194 | 0.50634 | 0.26171 |
| 4 | 2500 | 0.46226 | 0.239 | 0.29873 |
| 5 | 2500 | 0.44713 | 0.30396 | 0.24892 |
| 6 | 2500 | 0.21815 | 0.47151 | 0.31034 |
| 7 | 2500 | 0.12073 | 0.30907 | 0.5702 |
| 8 | 2500 | 0.4339 | 0.18723 | 0.37887 |
| 9 | 2500 | 0.25191 | 0.18695 | 0.56114 |
| 10 | 2500 | 0.08701 | 0.69609 | 0.2169 |
| 11 | 2500 | 0.31462 | 0.49271 | 0.19268 |
| 12 | 2500 | 0.40857 | 0.36551 | 0.22591 |
| 13 | 2500 | 0.20321 | 0.51516 | 0.28163 |
| 14 | 2500 | 0.31168 | 0.37399 | 0.31433 |
| 15 | 2500 | 0.33855 | 0.45624 | 0.2052 |
| 16 | 2500 | 0.40468 | 0.37178 | 0.22354 |

**Table S2**. A list of the stationary state averages resulting from the benchmark kMC simulations. Data is given to the same significant figures as was used in simulation for possible reproducibility.

| Simulation Number | $N_{sys}$ | $\sigma_{AA}^2$ | $\sigma_{BB}^2$ | $\sigma_{CC}^2$ | $\sigma_{AB}^2$ | $\sigma_{AC}^2$ | $\sigma_{BC}^2$ |
|---|---|---|---|---|---|---|---|
| 1 | 100 | 0.04713 | 0.03923 | 0.04977 | -6.41217·10⁻⁴ | -0.00158 | -8.97577·10⁻⁴ |
| 2 | 100 | 0.04884 | 0.04954 | 0.0197 | -0.00223 | $-1.59706 \cdot 10^{-4}$ | $-2.2852 \cdot 10^{-4}$ |
| 3 | 100 | 0.04918 | 0.04977 | 0.02924 | -0.00202 | $-3.98553 \cdot 10^{-4}$ | $-4.56595 \cdot 10^{-4}$ |
| 4 | 100 | 0.04964 | 0.04342 | 0.0401 | -0.00137 | -0.00109 | $-5.14728 \cdot 10^{-4}$ |
| 5 | 100 | 0.04698 | 0.04024 | 0.0498 | $-6.73463 \cdot 10^{-4}$ | -0.00153 | $-9.45753 \cdot 10^{-4}$ |
| 6 | 100 | 0.03179 | 0.05002 | 0.04905 | $-5.53246 \cdot 10^{-4}$ | $-4.57429 \cdot 10^{-4}$ | -0.00195 |
| 7 | 100 | 0.04544 | 0.04925 | 0.04586 | -0.00119 | $-8.71495 \cdot 10^{-4}$ | -0.00123 |
| 8 | 100 | 0.04916 | 0.03895 | 0.04903 | $-7.65082 \cdot 10^{-4}$ | -0.00165 | $-7.51944 \cdot 10^{-4}$ |
| 9 | 100 | 0.04819 | 0.04922 | 0.04101 | -0.00153 | $-7.90832 \cdot 10^{-4}$ | $-8.91152 \cdot 10^{-4}$ |
| 10 | 100 | 0.03482 | 0.04946 | 0.04516 | $-8.09765 \cdot 10^{-4}$ | $-4.02833 \cdot 10^{-4}$ | -0.00164 |
| 11 | 100 | 0.04976 | 0.04753 | 0.0402 | -0.00156 | $-9.1701 \cdot 10^{-4}$ | $-6.9941 \cdot 10^{-4}$ |
| 12 | 100 | 0.05013 | 0.0425 | 0.04432 | -0.00118 | -0.00134 | $-6.28549 \cdot 10^{-4}$ |
| 13 | 100 | 0.04444 | 0.04072 | 0.04994 | $-5.69518 \cdot 10^{-4}$ | -0.00141 | -0.00109 |
| 14 | 100 | 0.04538 | 0.04359 | 0.04973 | $-7.43372 \cdot 10^{-4}$ | -0.00132 | -0.00116 |
| 15 | 100 | 0.04991 | 0.03987 | 0.0472 | $-9.26511 \cdot 10^{-4}$ | -0.00156 | $-6.63485 \cdot 10^{-4}$ |
| 16 | 100 | 0.02842 | 0.05 | 0.04888 | $-4.58972 \cdot 10^{-4}$ | $-3.4856 \cdot 10^{-4}$ | -0.00204 |
| 1 | 256 | 0.03048 | 0.03104 | 0.01855 | $-7.74181 \cdot 10^{-4}$ | $-1.5461 \cdot 10^{-4}$ | $-1.89587 \cdot 10^{-4}$ |
| 2 | 256 | 0.03091 | 0.0287 | 0.0278 | $-5.03394 \cdot 10^{-4}$ | $-4.52213 \cdot 10^{-4}$ | $-3.20569 \cdot 10^{-4}$ |
| 3 | 256 | 0.03056 | 0.02207 | 0.03117 | $-2.24727 \cdot 10^{-4}$ | $-7.09348 \cdot 10^{-4}$ | $-2.62273 \cdot 10^{-4}$ |
| 4 | 256 | 0.01524 | 0.03091 | 0.03006 | $-1.42049 \cdot 10^{-4}$ | $-9.02465 \cdot 10^{-5}$ | $-8.13109 \cdot 10^{-4}$ |
| 5 | 256 | 0.02318 | 0.0278 | 0.03104 | $-1.73247 \cdot 10^{-4}$ | $-3.6395 \cdot 10^{-4}$ | $-5.99756 \cdot 10^{-4}$ |
| 6 | 256 | 0.03046 | 0.0299 | 0.02768 | $-5.27948 \cdot 10^{-4}$ | $-4.00129 \cdot 10^{-4}$ | $-3.66036 \cdot 10^{-4}$ |
| 7 | 256 | 0.02884 | 0.02743 | 0.03088 | $-3.15276 \cdot 10^{-4}$ | $-5.16466 \cdot 10^{-4}$ | $-4.37284 \cdot 10^{-4}$ |
| 8 | 256 | 0.0229 | 0.02192 | 0.02871 | $-9.0501 \cdot 10^{-5}$ | $-4.34121 \cdot 10^{-4}$ | $-3.89993 \cdot 10^{-4}$ |
| 9 | 256 | 0.03073 | 0.02389 | 0.02664 | $-4.02948 \cdot 10^{-4}$ | $-5.41548 \cdot 10^{-4}$ | $-1.68007 \cdot 10^{-4}$ |
| 10 | 256 | 0.02613 | 0.01764 | 0.02891 | $-7.89503 \cdot 10^{-5}$ | $-6.03753 \cdot 10^{-4}$ | $-2.3227 \cdot 10^{-4}$ |
| 11 | 256 | 0.02025 | 0.02974 | 0.03108 | $-1.64278 \cdot 10^{-4}$ | $-2.4566 \cdot 10^{-4}$ | $-7.20317 \cdot 10^{-4}$ |
| 12 | 256 | 0.03063 | 0.03103 | 0.01323 | $-8.63114 \cdot 10^{-4}$ | $-7.53856 \cdot 10^{-5}$ | $-9.96176 \cdot 10^{-5}$ |
| 13 | 256 | 0.03107 | 0.02517 | 0.02724 | $-4.28363 \cdot 10^{-4}$ | $-5.37066 \cdot 10^{-4}$ | $-2.04954 \cdot 10^{-4}$ |
| 14 | 256 | 0.031 | 0.03045 | 0.01452 | $-8.38703 \cdot 10^{-4}$ | $-1.22497 \cdot 10^{-4}$ | $-8.84477 \cdot 10^{-5}$ |
| 15 | 256 | 0.02966 | 0.03029 | 0.02782 | $-5.11487 \cdot 10^{-4}$ | $-3.6802 \cdot 10^{-4}$ | $-4.06059 \cdot 10^{-4}$ |
| 16 | 256 | 0.02984 | 0.02963 | 0.02922 | $-4.57085 \cdot 10^{-4}$ | $-4.33364 \cdot 10^{-4}$ | $-4.20692 \cdot 10^{-4}$ |
| 1 | 2500 | 0.00951 | 0.00983 | 0.00833 | $-5.88437 \cdot 10^{-5}$ | $-3.16201 \cdot 10^{-5}$ | $-3.7745 \cdot 10^{-5}$ |
| 2 | 2500 | 0.00992 | 0.00758 | 0.00959 | $-3.19079 \cdot 10^{-5}$ | $-6.64756 \cdot 10^{-5}$ | $-2.55269 \cdot 10^{-5}$ |
| 3 | 2500 | 0.00835 | 0.00987 | 0.00861 | $-4.64515 \cdot 10^{-5}$ | $-2.32098 \cdot 10^{-5}$ | $-5.09466 \cdot 10^{-5}$ |
| 4 | 2500 | 0.01009 | 0.00849 | 0.00912 | $-4.53699 \cdot 10^{-5}$ | $-5.64705 \cdot 10^{-5}$ | $-2.67353 \cdot 10^{-5}$ |
| 5 | 2500 | 0.00987 | 0.00908 | 0.00872 | $-5.19172 \cdot 10^{-5}$ | $-4.55156 \cdot 10^{-5}$ | $-3.05283 \cdot 10^{-5}$ |
| 6 | 2500 | 0.00828 | 0.00998 | 0.00924 | $-4.13491 \cdot 10^{-5}$ | $-2.71422 \cdot 10^{-5}$ | $-5.81909 \cdot 10^{-5}$ |
| 7 | 2500 | 0.00658 | 0.00917 | 0.0098 | $-1.56486 \cdot 10^{-5}$ | $-2.76005 \cdot 10^{-5}$ | $-6.84394 \cdot 10^{-5}$ |
| 8 | 2500 | 0.01002 | 0.0078 | 0.00978 | $-3.2786 \cdot 10^{-5}$ | $-6.76649 \cdot 10^{-5}$ | $-2.7982 \cdot 10^{-5}$ |
| 9 | 2500 | 0.00857 | 0.00778 | 0.00984 | $-1.86286 \cdot 10^{-5}$ | $-5.48987 \cdot 10^{-5}$ | $-4.19067 \cdot 10^{-5}$ |
| 10 | 2500 | 0.00557 | 0.00911 | 0.00821 | $-2.32866 \cdot 10^{-5}$ | $-7.74831 \cdot 10^{-6}$ | $-5.96932 \cdot 10^{-5}$ |
| 11 | 2500 | 0.00922 | 0.00986 | 0.00786 | $-6.01903 \cdot 10^{-5}$ | $-2.48447 \cdot 10^{-5}$ | $-3.69514 \cdot 10^{-5}$ |
| 12 | 2500 | 0.00981 | 0.00989 | 0.00841 | -6.16629·10⁻⁵ | -3.45697·10⁻⁵ | -3.60801·10⁻⁵ |
| 13 | 2500 | 0.00792 | 0.00992 | 0.00897 | -4.02971·10⁻⁵ | -2.24465·10⁻⁵ | -5.80892·10⁻⁵ |
| 14 | 2500 | 0.00925 | 0.00955 | 0.00911 | -4.68762·10⁻⁵ | -3.87458·10⁻⁵ | -4.43262·10⁻⁵ |
| 15 | 2500 | 0.00956 | 0.01001 | 0.00816 | -6.25967·10⁻⁵ | -2.88723·10⁻⁵ | -3.76847·10⁻⁵ |
| 16 | 2500 | 0.00996 | 0.00972 | 0.00831 | -6.22798·10⁻⁵ | -3.69315·10⁻⁵ | -3.21827·10⁻⁵ |

**Table S3**. A list of the stationary state variances and covariances resulting from the benchmark kMC simulations. Data is given to the same significant figures as was used in simulation for possible reproducibility.

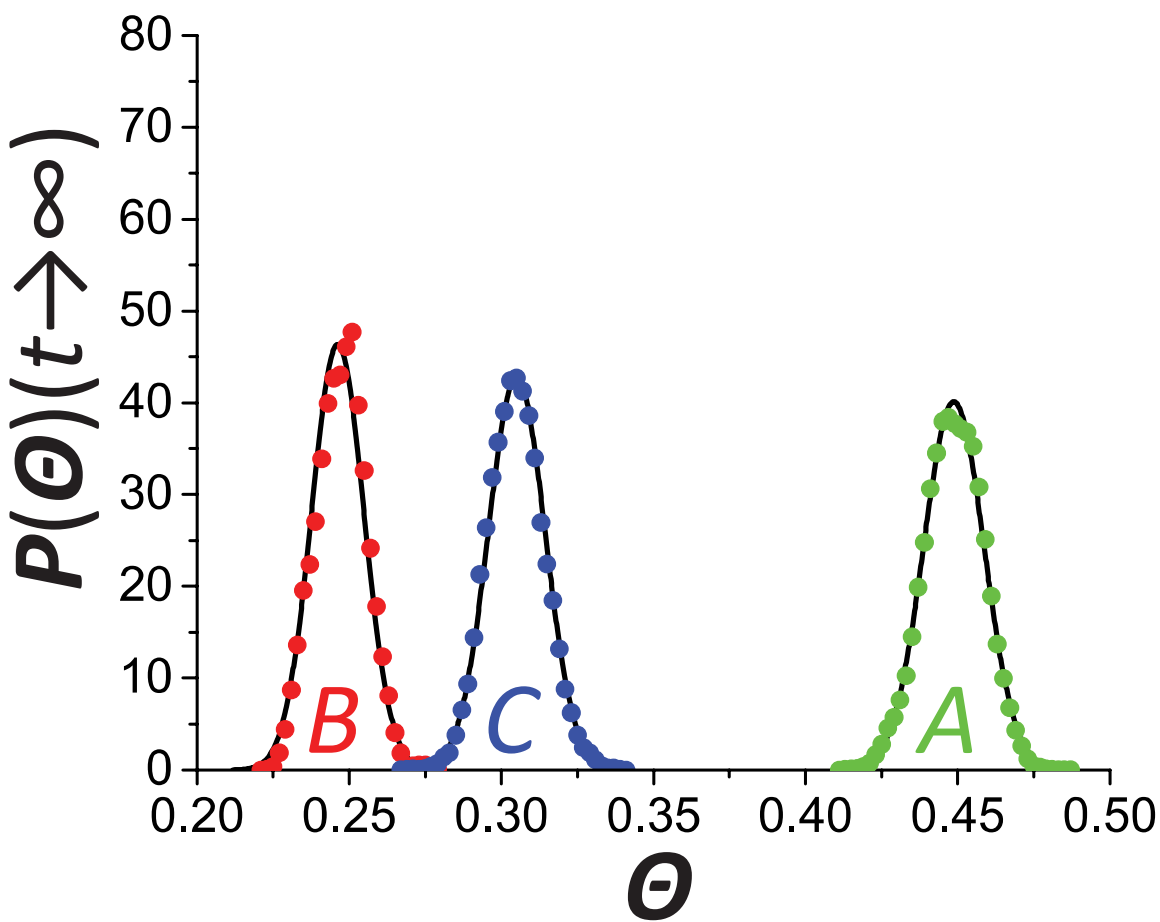

**Figure S1**. Comparisons of the MPT predicted variance and covariance in a Gaussian model against kMC numerical. The colored data is kMC and the solid black line an MPT prediction, where the kMC data was $(N_{sys}, k_{rxn}^{A \to B}, k_{rxn}^{B \to A}, k_{rxn}^{A \to C}, k_{rxn}^{C \to A}, k_{rxn}^{B \to C}, k_{rxn}^{C \to B})$=(100, 0.09997, 0.56300, 0.11552, 0.38781, 0.88192, 0.25927).

**Definition of variables.** A lot of variables are used in this manuscript, as an aid those variables and their definitions are given in Table S1.

| Variable | Definition |
|---|---|
| | |
| $\Sigma$ | Summation operator. For a series of numbers, $a_{i_1}$, $i_1$ from 1 to $n$, the summation operator is used to represent the sum of that series as $\sum_{i_1=1}^{n} a_{i_1} = a_1 + \cdots + a_n$. |
| $\Pi$ | Product operator. For a series of numbers, $a_{i_1}$, $i_1$ from 1 to $n$, the product operator is used to represent the product of that series as $\prod_{i_1=1}^{n} a_{i_1} = a_1 * \ldots * a_n$. |
| ! | The factorial symbol |
| $f$ | The probability mass function |
| $n_{sample}$ | The sample size of a statistical sample |
| $x_i$ | The number of states $i$ in a statistical sample |
| $E[X]$ | The expected value of some variable, $X$ |
| $Var[x_i]$ or $\sigma_{ii}^2$ | The variance of $x_i$ |
| $Covar[x_i, x_j]$ or $\sigma_{ij}^2$ | The covariance of $x_i$ with $x_j$ |
| $\Sigma$ | The covariance matrix |
| $a$ | A distance parameter corresponding the separation of lattice sites in cMRT |
| $\mathbf{k}^{(m)}$ | The rate constant hypermatrix for reactions of order $m$ in cMRT |
| $\boldsymbol{\theta}$ | Concentration vector containing concentrations of each species $i$ in cMRT |
| $\dot{\boldsymbol{\theta}}$ | Time derivative of concentration vector containing concentrations derivative of each species $i$ as $\dot{\theta}_i$ in cMRT |
| $i$ or $j$ | A state in either the MPT of physical systems |
| $p_i$ | The probability drawing a state $i$ from an MPT sample |
| $N_{sys}$ | The total number of states in either a canonical or grand canonical sysytem |
| $\{x_i\}$ | Some set of states of the type $x_i$ |
| $W\{x_i\}$ | The probability of the set of states $\{x_i\}$ |
| $w_i$ | The probability weight factor of an independent subsystem in either the canonical or grand canonical ensembles |
| $r_l$ | The rate of some reaction $l$ |
| $P_l$ | The probability of selection a reaction $l$ in kinetic Monte Carlo |
| $\Delta t$ | The discrete time step in kinetic Monte Carlo |
| $R$ | A random number between zero and one |

**Table S4.** Variable definitions. This variable list applies to the variables in the body of the manuscript.

# References

1. Eyring, H. The activated complex in chemical reactions. *J. Chem. Phys.* **3**, 9 (1935).
2. Laidler, K. J. & King, M. C. Development of transition-state theory. *The Journal of Physical Chemistry* **87**, 2657-2664, doi:10.1021/j100238a002 (1983).
3. Vineyard, G. H. Frequency Factors and Isotope Effects in Solid State and Rate Processes. *J. Phys. Chem. Solids* **3**, 121-127 (1957).
4. Medford, A. J. *et al.* CatMAP: A Software Package for Descriptor-Based Microkinetic Mapping of Catalytic Trends. *Catal. Lett.* **145**, 794-807, doi:10.1007/s10562-015-1495-6 (2015).
5. Norskov, J. K., Bligaard, T., Rossmeisl, J. & Christensen, C. H. Towards the computational design of solid catalysts. *Nature Chemistry* **1**, 37-46 (2009).
6. Francis, M. F. & Curtin, W. A. Mechanical work makes important contributions to surface chemistry at steps. *Nature Communications* **6**, 1-7, doi:10.1038/ncomms7261 (2015).
7. Francis, M. F. & Curtin, W. A. Elastic Effects in Adsorbate–Adsorbate Interactions of C and S on a Stepped Ru Surface. *The Journal of Physical Chemistry C* **121**, 16761-16769, doi:10.1021/acs.jpcc.7b02422 (2017).
8. Francis, M. F. & Curtin, W. A. Mechanical Stress Combined with Alloying May Allow Continuous Control Over Reactivity: Strain Effects on CO Dissociation and Subsequent Methanation Catalysis over Ni(211), Ni3Fe(211), and NiFe(112). *The Journal of Physical Chemistry C* **121**, 6113-6119, doi:10.1021/acs.jpcc.6b12329 (2017).
9. Taylor, C. D. A First-Principles Surface Reaction Kinetic Model for Hydrogen Evolution under Cathodic and Anodic Conditions on Magnesium. *J. Electrochem. Soc.* **163**, C602-C608, doi:10.1149/2.1171609jes (2016).
10. Lillard, R. S., Wang, G. F. & Baskes, M. I. The Role of Metallic Bonding in the Crystallographic Pitting of Magnesium. *J. Electrochem. Soc.* **153**, B358-B364, doi:10.1149/1.2218108 (2006).
11. Francis, M. F. & Taylor, C. D. First-principles insights into the structure of the incipient magnesium oxide and its instability to decomposition: Oxygen chemisorption to Mg(0001) and thermodynamic stability. *PhRvB* **87**, 075450, doi:10.1103/PhysRevB.87.075450 (2013).
12. Taylor, C. D. & Liu, X.-Y. Investigation of structure and composition control over active dissolution of Fe–Tc binary metallic waste forms by off-lattice kinetic Monte Carlo simulation. *J. Nucl. Mater.* **434**, 382-388, doi:10.1016/j.jnucmat.2012.11.039 (2013).
13. Balwierz, P. J. *et al.* ISMARA: Automated modeling of genomic signals as a democracy of regulatory motifs. *Genome Research*, doi:10.1101/gr.169508.113 (2014).
14. Lück, S., Thurley, K., Thaben, Paul F. & Westermark, Pål O. Rhythmic Degradation Explains and Unifies Circadian Transcriptome and Proteome Data. *Cell Reports* **9**, 741-751, doi:10.1016/j.celrep.2014.09.021.
15. Pathria, R. K. in *Statistical Mechanics (Second Edition)*    90-103 (Butterworth-Heinemann, 1996).
16. Pathria, R. K. in *Statistical Mechanics (Second Edition)*    43-89 (Butterworth-Heinemann, 1996).
17. Francis, M. F. & Taylor, C. D. in *Electrochemical Society*    (Toronto, 2013).

18    Schwoebel, R. L. Oxide Formation on Magnesium Single Crystals. II. Structure and Orientation. *J. Appl. Phys.* **34**, 2784-2788 (1963).
19    Schwoebel, R. L. Oxide Formation on Magnesium Single Crystals. I. Kinetics of Growth. *J. Appl. Phys.* **34**, 2776-2783 (1963).
20    Bligaard, T. *et al.* The Brønsted–Evans–Polanyi relation and the volcano curve in heterogeneous catalysis. *J. Catal.* **224**, 206-217 (2004).
21    Varvenne, C., Luque, A. & Curtin, W. A. Theory of strengthening in fcc high entropy alloys. *Acta Mater.* **118**, 164-176, doi:10.1016/j.actamat.2016.07.040 (2016).
22    Christensen, C. H. & Nørskov, J. K. A molecular view of heterogeneous catalysis. *The Journal of Chemical Physics* **128**, 182503, doi:http://dx.doi.org/10.1063/1.2839299 (2008).
23    Honkala, K. *et al.* Ammonia Synthesis from First-Principles Calculations. *Science* **307**, 555-558, doi:10.1126/science.1106435 (2005).
24    Francis, M. F. Continuum Microkinetic Rate Theory of Lattice Systems: Formalization, Current Limitations, and a Possible Basis for Continuum Rate Theory. *The Journal of Physical Chemistry A* **122**, 7267-7275, doi:10.1021/acs.jpca.8b06238 (2018).
25    Francis, M. F. Correction to "Continuum Microkinetic Rate Theory of Lattice Systems: Formalization, Current Limitations, and a Possible Basis for Continuum Rate Theory". *The Journal of Physical Chemistry A* **122**, 9149-9149, doi:10.1021/acs.jpca.8b10093 (2018).
26    Bortz, A. B., Kalos, M. H. & Lebowitz, J. L. A New Algorithm for Monte Carlo Simulation of Ising Spin Systems. *J. Comput. Phys.*, 10/18 (1975).
27    Gillespie, D. T. A general method for numerically simulating the stochastic time evolution of coupled chemical reactions. *J. Comput. Phys.* **22**, 403-434, doi:https://doi.org/10.1016/0021-9991(76)90041-3 (1976).
28    Kurtz, T. G. The Relationship between Stochastic and Deterministic Models for Chemical Reactions. *The Journal of Chemical Physics* **57**, 2976-2978, doi:10.1063/1.1678692 (1972).
29    Gameiro, M., Mischaikow, K. & Wanner, T. Evolution of pattern complexity in the Cahn–Hilliard theory of phase separation. *Acta Mater.* **53**, 693-704, doi:https://doi.org/10.1016/j.actamat.2004.10.022 (2005).
30    Cook, H. E. Brownian motion in spinodal decomposition. *Acta Metall.* **18**, 297-306, doi:https://doi.org/10.1016/0001-6160(70)90144-6 (1970).
31    Cahn, J. W. & Hilliard, J. E. Free Energy of a Nonuniform System. I. Interfacial Free Energy. *The Journal of Chemical Physics* **28**, 258-267, doi:10.1063/1.1744102 (1958).